\documentclass[prb,twocolumn,superscriptaddress,floatfix]{revtex4}
\usepackage{graphicx}
\newcommand{\pdag}{{\phantom{\dagger}}}

\begin{document}
\title{Semi--analytical solution of the Kondo model in a magnetic field}
\author{C.\ Slezak}
\affiliation{Center for electronic correlations and magnetism,
Theoretical Physics III, Institute for Physics, University of Augsburg,
86135 Augsburg, Germany}
\affiliation{Department of Physics, University of Cincinnati, Cincinnati
OH 45221, USA}
\author{S.\ Kehrein}
\affiliation{Center for electronic correlations and magnetism,
Theoretical Physics III, Institute for Physics, University of Augsburg,
86135 Augsburg, Germany}
\author{Th.\ Pruschke}
\affiliation{Center for electronic correlations and magnetism,
Theoretical Physics III, Institute for Physics, University of Augsburg,
86135 Augsburg, Germany}
\author{M.\ Jarrell}
\affiliation{Department of Physics, University of Cincinnati, Cincinnati
OH 45221, USA}

\begin{abstract}
The single impurity Kondo model at zero temperature in a magnetic field is
solved by a
semi--analytical approach based on the flow equation method. The resulting
problem is shown to be equivalent to a resonant level
model with a non--constant hybridization function. This nontrivial {\em effective
hybridization function} encodes the quasiparticle interaction
in the Kondo limit, while the magnetic field enters as the impurity orbital
energy. The evaluation of static and dynamic quantities of the
strong--coupling Kondo model becomes very simple in this effective model.
We present results for thermodynamic quantities and the dynamical
spin--structure factor and compare them with NRG calculations.
\end{abstract}
\date{\today}
\maketitle

\section{Introduction}
The single impurity Kondo or $s$--$d$ model (SIKM) \cite{kondo}
\begin{equation}\label{equ:SIKM}
H_{\rm SIKM} = \sum\limits_{\vec k\sigma}\epsilon^\pdag_{\vec k}
c^\dagger_{\vec k\sigma}c^\pdag_{\vec k\sigma}
+J\sum\limits_{\vec k\vec k'}\sum\limits_{\alpha\beta}
c^\dagger_{\vec k\alpha}\, {\vec S}\cdot\vec \sigma^\pdag_{\alpha\beta}\,
c^\pdag_{\vec k'\beta}
\end{equation}
is, together with its close relative, the single impurity Anderson model
(SIAM) \cite{anderson}, one of the
fundamental models in the theory of correlated electron systems.
It has been studied extensively over the past four decades \cite{hewson}, but
despite its simplicity, no complete analytic solution exists that provides
information on both thermodynamic and dynamical quantities. For example,
the Bethe ansatz solution \cite{bethe,hewson} has solved all
universal properties of the Kondo problem including its high-- to
low--temperature crossover behavior. But the Bethe ansatz
requires the limit of infinite conduction band width
and cannot be used to calculate dynamical quantities beyond the low--energy limit.

Thus, numerical methods, like Wilson's numerical renormalization group (NRG)
\cite{Wilson,hewson} or quantum Monte--Carlo (QMC) \cite{fye}, in connection
with maximum--entropy methods \cite{MEM}
have to be employed to access dynamical quantities on all energy scales.
In both methods, evaluation of dynamical properties, and quantities
related to them, like the Korringa--Shiba relation or the Friedel sum rule
\cite{hewson},
suffer from unavoidable numerical errors. QMC simulations, in addition, cannot be
used at zero temperature and are restricted to comparatively large values of
$J$.\cite{fye} Moreover, a reliable evaluation of
single-- and two--particle spectra and related quantities in an external magnetic
field, as well as their comparison and interpretation within a local Fermi liquid
picture \cite{hewson} become rather problematic \cite{costi,logan4}, especially
in the limit of vanishing external magnetic field.
Approximate analytical techniques, like perturbation expansions or
$1/N$--expansions \cite{hewson}, typically only describe certain
properties of the Kondo model correctly.
A notable exception in this respect is the so--called local moment approach
(LMA) \cite{logan}. This perturbative approach is very accurate in most cases \cite{logan2},
including those of nonvanishing external magnetic fields.\cite{logan3,logan4}

Besides its relevance for the study of moment formation in metals, the SIKM
(\ref{equ:SIKM}) has gained new importance as
input for investigating non--dilute correlated electron systems like heavy
fermion materials within dynamical mean--field theory (DMFT).\cite{DMFT}
Here a reliable method for calculating single--particle
correlation functions, especially close to the Fermi energy, is extremely
important.

In this paper, we propose a new {\em non--perturbative} semi--analytical approach to the Kondo problem
based on Wegner's flow equation method \cite{wegner} and previous work on
applications of flow equations to strong--coupling problems
\cite{strong_coupling,Hofstetter_Kehrein}. We will show that to a very good approximation,
many physical quantities of the Kondo model can be calculated from a
resonant level model (RLM), where the interacting features
of the Kondo model are encoded in a non--constant {\em effective
hybridization
function} of this resonant level model. Surprisingly, this noninteracting
effective model describes both universal low--energy properties like
the Wilson ratio as well as high--energy
power laws and logarithmic corrections with very good accuracy. Due to
the noninteracting nature of this effective model this mapping allows immediate
insights into the physics of the SIKM, for example the dependence of its static
and dynamical quantities on a local magnetic field.

After presenting our approach in the next section, we will discuss several static
quantities at $T=0$ as a function of a local magnetic field
and derive analytical expressions
for their asymptotic behavior. As an example for a dynamical quantity, we will
then discuss the spin--structure factor and the Korringa--Shiba relation.
An outlook on potential future applications of our approach
concludes this paper.

\section{Mapping to a resonant level model}
\subsection{Principles of the flow-equation method}
The general framework of the flow equation method\cite{wegner} and its
application to the Kondo model has been explained in detail in
Ref.~\onlinecite{Hofstetter_Kehrein}.
Here we will only repeat the main steps in order to make this paper self--contained,
and refer to Ref.~\onlinecite{Hofstetter_Kehrein} for more details.

The key idea of the flow equation approach consists in performing a continuous
sequence of infinitesimal unitary transformations on a given Hamiltonian
\begin{equation}
\frac{dH(B)}{dB}=[\eta(B),H(B)] \ .
\label{floweq_1}
\end{equation}
With an anti--Hermitian generator $\eta(B)$ the solution of equation
(\ref{floweq_1}) describes a family of unitarily equivalent Hamiltonians
$H(B)$ parameterized by the {\em flow parameter}~$B$. By choosing
$\eta(B)$ appropriately\cite{wegner} one can set up a framework that
diagonalizes a many--particle Hamiltonian $H(B=0)$, i.e.
$H(B=\infty)$ becomes diagonal.

The concrete realization of this
approach for the Kondo model was discussed in Ref.~\onlinecite{Hofstetter_Kehrein}.
The starting point is the bosonized form\cite{boson} of the
Hamiltonian~(\ref{equ:SIKM}).
Since we will be mainly interested in describing the basic ideas of our
approach, we restrict ourselves to a linear dispersion relation.
Notice, however, that the flow equation approach can also be used
for a nontrivial conduction band density of states as it does not rely
on the integrability of the model.\cite{perspective}
With a linear dispersion relation
the Kondo problem becomes effectively one--dimensional,
the charge density excitations in (\ref{equ:SIKM}) decouple, and we only need
to look at the spin density part
\begin{equation}
H=H_0
 - \frac{J}{\sqrt{8 \pi^2}} \partial_x \Phi(0)\, S^z
 + \frac{J}{4\pi a}\, \left(e^{i \sqrt{2} \Phi(0)}\,
S^- + {\rm h.c.} \right)
\label{bosonic_Kondo}
\end{equation}
with $H_0=\sum_{q>0}q\, \sigma(q)\, \sigma(-q)$.
Here
$\sigma(p) = \frac{1}{\sqrt{2|p|}} \sum_{q} (c^\dagger_{p+q \uparrow}
c^\pdag_{q \uparrow} - c^\dagger_{p+q \downarrow} c^\pdag_{q
\downarrow} )$
are the bosonic spin density modes with the bosonic spin density field defined by
$\Phi(x) = -i \sum_{q \ne 0} \frac{\sqrt{|q|}}{ q}\:
e^{-i q x - a|q|/2}\: \sigma(q)$. For simplicity we have set the
Fermi velocity $v_F=1$. $a$~is proportional to the inverse conduction
band width. All our latter results will be expressed as universal
functions of the low--energy Kondo scale~$T_{\rm K}$, and we can consider
(\ref{bosonic_Kondo}) to be equivalent to our original Kondo Hamiltonian
if $T_{\rm K}\ll a^{-1}$.

Eq.~(\ref{bosonic_Kondo}) was used as the starting point $H(B=0)$
of the flow equation approach in Ref.~\onlinecite{Hofstetter_Kehrein}. Away
from the Toulouse point the unitary equivalence of the flow holds only
approximately, but this approximation can be controlled by a small
parameter\cite{strong_coupling} and yields very accurate
results. During the flow the Hamiltonian can be parameterized as
\begin{eqnarray}
H(B)&=&H_0+\sum_p g_p(B)\, \left(C^\dagger_p(\lambda(B))\, S^- +
{\rm h.c.} \right)
\label{H(B)} \\
&&+\sum_p \omega_{p}(B)\,
\left[C^\dagger_p(\lambda(B_p)), C^\pdag_p(\lambda(B_p)) \right] \ . \nonumber
\end{eqnarray}
Here $B_p\stackrel{\rm def}{=}p^{-2}$, and
$C^\dagger_p(\lambda),~C^\pdag_p(\lambda)$ denote normalized vertex operators
with scaling dimension~$\lambda>0$ in momentum space,
\begin{eqnarray}
C^\dagger_p(\lambda) &\stackrel{\rm def}{=}&
\left(\frac{\Gamma(\lambda^2)}{2 \pi a\,L}\right)^{1/2}
|p a|^{(1-\lambda^2)/2}
\int_{-L/2}^{L/2} dx\,e^{ipx+i \lambda \Phi(x)} \nonumber \\
C^\pdag_p(\lambda) &\stackrel{\rm def}{=}&
\left(\frac{\Gamma(\lambda^2)}{2 \pi a\,L}\right)^{1/2}
|p a|^{(1-\lambda^2)/2}
\int_{-L/2}^{L/2} dx\,e^{-ipx-i \lambda \Phi(x)} \nonumber
\end{eqnarray}
that obey
$\langle\Omega|
C^\pdag_p(\lambda)\,C^\dagger_{p'}(\lambda)|\Omega\rangle=\delta_{pp'}\theta(p)$
and
$\langle\Omega|
C^\dagger_p(\lambda)\,C^\pdag_{p'}(\lambda)|\Omega\rangle=\delta_{pp'}\theta(-p)$.
For the special case $\lambda=1$ they fulfill fermionic anticommutation relations
$\{C^\dagger_p(1),C^\pdag_{p'}(1)\}=\delta_{pp'}$ and can therefore be interpreted
as creation and annihilation operators for fermions.

In Ref.~\onlinecite{Hofstetter_Kehrein} the following flow equations for the
parameters in (\ref{H(B)}) have been derived
\begin{eqnarray}
\frac{dg_p}{dB}&=&
-p^2 g_p+\frac{2\pi}{\Gamma(\lambda^2)}\sum_{q\neq p} \frac{p+q}{p-q}\: g_p\, g_q^2
\: |qa|^{\lambda^2-1} \nonumber \\
&&+\frac{1}{4}\:g_p\:\ln(B/a^2)\:\frac{d\lambda^2}{dB}
\label{feq_1} \\
\frac{d\omega_q}{dB}&=&\frac{2\pi}{\Gamma(\lambda^2)} \:q\, g_q^2\:
|qa|^{\lambda^2-1}
\label{feq_2}
\end{eqnarray}
and a differential equation for the flow of the scaling dimension
\begin{equation}
\frac{d\lambda^2}{dB}=\frac{8\pi\lambda^2 (1-\lambda^2)}{\Gamma(\lambda^2)}
\sum_q g_q\, g_{-q}\: |qa|^{\lambda^2-1} \ .
\label{feq_3}
\end{equation}
It can be shown\cite{Hofstetter_Kehrein} that
one always finds $\lambda\stackrel{B\rightarrow\infty}{\longrightarrow} 1$
in the strong--coupling phase of the Kondo model,
i.e.\ in the low--energy limit the vertex operators in (\ref{H(B)})
become fermions. In the following we will use an improved version of
the above flow equations by taking into account that all approximations
should be performed with respect to the interacting ground state:
It turns out that the only necessary modification in (\ref{feq_1}), (\ref{feq_2})
and (\ref{feq_3}) is that the exponent in $|qa|^{\lambda^2-1}$
gets replaced by $\lambda^2(B_q)-1$, i.e.\ it is not a running exponent
anymore.\cite{Kehrein_tobepublished}

\subsection{Equivalence to a resonant level model}
Now we will compare this system of differential equations with the flow
equations for a resonant level model (RLM). This will lead
to the {\em key result} of this paper: the RLM can
be used as an {\em effective model} for the complicated strong--coupling Kondo
model.  

The Hamiltonian of the resonant level model is 
given by
\begin{equation}\label{equ:SIAM}
H_{\rm RLM} = \sum\limits_{k}\epsilon^\pdag_{k}
c^\dagger_{k}c^\pdag_{k}+\epsilon^\pdag_{d}d^\dagger d+
\sum\limits_{k}V^\pdag_{k}(c^\dagger_{k} d+d^\dagger c^\pdag_{k})
\label{equ:RML}
\end{equation}
Following the same flow equation approach as previously in the SIKM, we
establish a solution to the RLM (\ref{equ:SIAM}). A detailed description of
the flow equation solution
can be found in  Ref.~\onlinecite{SIAM}. One finds the following flow equations
for the parameters in (\ref{equ:RML})
\begin{eqnarray}
\frac{dV_k}{dB}&=&
-V_k (\epsilon_d-\epsilon_k)^2+ \sum\limits_{q\neq k}V_k^\pdag
V_q^2\,\frac{\epsilon_k+\epsilon_q-2\epsilon_d}{\epsilon_k-\epsilon_q}
\label{feq_a1} \\
\frac{d\epsilon_d}{dB}&=&-2 \sum\limits_{k}V_k^2\,(\epsilon_k-\epsilon_d)
\label{feq_a2} \\
\frac{d\epsilon_k}{dB}&=&2V_k^2(\epsilon_k-\epsilon_d)
\label{feq_a3} 
\end{eqnarray}

It should be noted that this yields the exact
analytical solution. Having established the flow equations to solve
both the SIKM and the 
RLM, respectively, one can now show an approximate equivalence of these two
models. We introduce the substitution
\begin{eqnarray}
V_k^2=\frac{2\pi}{\Gamma(\lambda^2(B_k))}\,g_k^2|k a|^{\lambda^2(B_k)-1}
\label{subst}
\end{eqnarray}
and notice that with this substitution the two set of flow equations
(\ref{feq_1},\ref{feq_2}) and (\ref{feq_a1}--\ref{feq_a3}) become equivalent for
$\epsilon_d=0$ in the RLM, with the exception of
the logarithmic term in (\ref{feq_1}). 
Thus we now have established an approximate mapping of the SIKM onto the RLM
by means of (\ref{subst}) in the sense that their flow equation
diagonalization is identical.
We shall refer to this relation by introducing the 
{\em effective hybridization function} $\Delta_{\rm
  eff}(\epsilon)=\pi\sum_{k}V^2_k\delta(\epsilon-\epsilon_k)$:
the RLM with this non--trivial hybridization function can be used
as an effective model for the SIKM in the Kondo limit (small coupling
limit) $\rho_0 J\rightarrow 0$.
Since this noninteracting RLM is a simple, quadratic Hamiltonian, this mapping
will allow us to read off and understand many properties of the complicated
many--body Kondo physics in an intuitive and straightforward way. 
{\em It will turn out that the deviations of $\Delta_{\rm eff}(\epsilon)$ from
a constant hybridization function encode the quasiparticle interaction 
and therefore the many--body Kondo physics in this quadratic effective
Hamiltonian.}

Notice that the above mapping between the SIKM and the RLM becomes {\em exact}
at the Toulouse point \cite{Toulouse} 
$\rho_0 J=2\pi(2-\sqrt{2})$ since $\lambda^2(B)=1$
for all flow parameters~$B$. One easily verifies that the effective RLM
then has a constant hybridization function, 
$\Delta_{\rm eff}(\epsilon)=\frac{\pi}{4}\rho_0 J^2$.
In this case, our mapping just reduces to the observation already made
by Toulouse that the partition function of the Kondo model for
this specific coupling constant~$J$ is exactly
equivalent to the partition function of a quadratic Hamiltonian.\cite{Toulouse}

In order to specify the function $\Delta_{\rm eff}(\epsilon)$ in the Kondo limit
it is best to not directly use relation (\ref{subst}), but to determine
the effective hybridization function from matching a correlation
function in the SIKM and the RLM. We have chosen the 
$\langle S^+(t) S^-(0) \rangle$--correlation
function, evaluated it with respect to (\ref{H(B)}) for $B=0$, and 
then chose $\Delta_{\rm eff}(\epsilon)$ in the RLM such that this
coincided with the $\langle d^\dagger(t) d(0) \rangle$--correlation function.
The resulting $\Delta_{\rm eff}(\epsilon)$ agrees with (\ref{subst})
in the high-- and low--energy regimes, with deviations only in the
crossover region. However, the mapping from the Kondo model to the
effective RLM becomes better since this procedure manages to
partly also take the logarithmic term in (\ref{feq_1}) into account.

\begin{figure}[htb]
\begin{center}
\includegraphics[width=0.45\textwidth,clip]{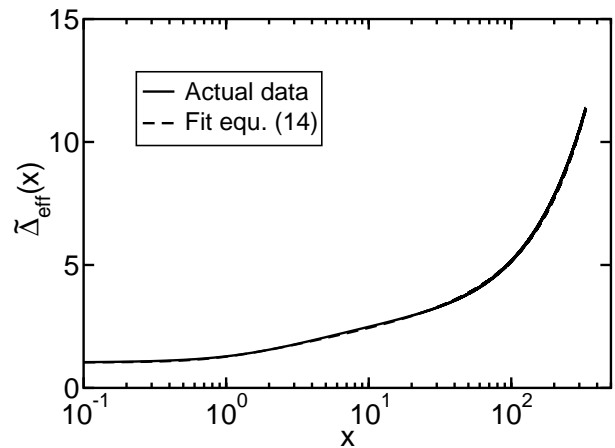}
\end{center}
\caption[]{The dimensionless effective hybridization function
$\tilde\Delta_{\rm eff}(x)$ evaluated for $\rho_0 J=0.1$
in the SIKM. The resulting function is symmetric and is only plotted for
$x>0$. It coincides with the universal form for $x\lesssim 30$.
The dashed line represents the fit
(\ref{equ:fit}) and is nearly indistinguishable from the actual data.
Notice especially the appearance of logarithmic behavior in the
crossover region.\label{fig:Fig1}}
\end{figure}

The resulting effective hybridization function can
be scaled into a dimensionless form with one dimensionful parameter
$\Delta_{\rm eff}^0 \propto T_{\rm K}$
\begin{equation}
\Delta_{\rm eff}(\epsilon)=\Delta_{\rm eff}^0\:
\tilde\Delta_{\rm eff}(\epsilon/\Delta_{\rm eff}^0) \ .
\end{equation}
$\tilde\Delta_{\rm eff}(x)$ is a universal function in the 
Kondo limit ($\rho_0 J\rightarrow 0$).
It is depicted in Fig.~\ref{fig:Fig1} for $\rho_0 J=0.1$,
and coincides with its universal form for $|x|\lesssim 30$
($|\epsilon|\lesssim 60T_{\rm K}$, i.e. this should be sufficient
for most practical purposes\cite{numerics}). For larger energies
the effective hybridization function begins to cross over
into linear behavior with logarithmic corrections
depending on the {\em bare} coupling~$\rho_0 J$.
% Fit fuer effektive Hybridisierungsfunktion
The following function provides an excellent fit
(see Fig.~\ref{fig:Fig1})
\begin{equation}
  \label{equ:fit}
  \tilde\Delta_{\rm eff}(x)=
\begin{array}[t]{l}
\displaystyle 1+
\frac{1}{2}\,a_1\ln\left(1+\left(\frac{x}{a_0}\right)^2\right)\\[5mm]
\displaystyle
+a_2\left(\arctan\left|\frac{x}{a_0}\right|
-\left|\frac{x}{a_0}\right|\right)\left(
1-\ln\left|\frac{x}{a_0}\right|\right)
\end{array}
\end{equation}
with the parameters from Table~\ref{tab:fit}.
\begin{table}[htb]
\begin{center}
\begin{tabular}{|c|c|c|}
\hline
$a_0$ & $a_1$ &$a_2$\\
\hline
~~~0.829~~~ & ~~~0.536~~~ & ~~~0.00324~~~ \\
\hline
\end{tabular}
\end{center}
\caption[]{Result of the fit (\ref{equ:fit}) to the effective hybridization
$\tilde\Delta_{\rm eff}(x)$.\label{tab:fit}}
\end{table}

A similar analysis based on the comparison of flow equations shows that
the above mapping between the SIKM and the noninteracting RLM can be
extended to the case of a Kondo Hamiltonian (\ref{equ:SIKM}) with a
nonvanishing local magnetic field~$h$
\begin{equation}
H_{\rm SIKM}+g\mu_B\,h\,S_z
\end{equation}
by setting $\epsilon_d=g\mu_B h$ in the RLM.
However, the mapping with the above effective hybridization function becomes
less accurate for $g\mu_B |h|\gtrsim T_{\rm K}$ due to the approximate nature of the
flow equation solution (\ref{feq_1}--\ref{feq_3}). We will discuss this
point in more detail below.

Summing up, as long as we are interested in static quantities in
a local magnetic field smaller than approximately~$T_{\rm K}$ and/or dynamical
correlation functions for energies smaller than approximately~$60T_{\rm K}$,
we can use the RLM with the effective hybridization function (\ref{equ:fit})
to describe the physics of the SIKM in the small coupling limit.
The only undetermined parameter in the RLM is the energy scale
$\Delta^0_{\rm eff}$ that explicitly depends on~$J$. This overall
energy scale is proportional to $T_{\rm K}$. Notice that the
non--perturbative behavior of this energy scale
\begin{equation}
T_{\rm K} \propto e^{-1/\rho_0 J}
\end{equation}
follows correctly from the original flow equations (\ref{feq_1}--\ref{feq_3}),
compare Ref.~\onlinecite{Hofstetter_Kehrein}.

\subsection{Calculation of physical quantities}
Once the mapping between the SIKM (\ref{equ:SIKM}) and the effective
RLM (\ref{equ:SIAM})
has been established, one can readily calculate
physical quantities for the Kondo model. One complication arises from the fact
that operators of the
original SIKM have to be transformed by a unitary transformation analogous to
(\ref{floweq_1}). In the
language of the effective resonant level model they will thus in general
correspond to more complicated
many--particle operators. Since the intention of this paper is to demonstrate the
potential of our mapping in a pedagogical setting, we will
concentrate on two quantities that remain simple under these
transformations:
i)~the $z$-component of the spin operator $S_z$, which becomes
$S_z = d^\dagger d-1/2$, and ii)~the Hamiltonian itself. 

From the latter we obtain the
internal energy $U_{\rm imp}=\langle H-H_0\rangle$ and the Sommerfeld coefficient,
$\gamma_{\rm imp}(h)$. 
A straightforward calculation in the noninteracting RLM yields
\begin{equation}
\label{equ:gamma}
\gamma_{\rm imp}(h)=\frac{\pi^2k_{\rm B}^2}{3}\rho_{d}(0)\left(1-\Lambda'(0)\right)
\end{equation}
where $\Lambda'(\omega)$ denotes the derivative of
$$
\Lambda(\omega)=\frac{1}{\pi}{\rm P}\int d\epsilon\:
\frac{\Delta_{\rm eff}(\epsilon)}{\omega-\epsilon}
$$
and ${\rm P}\int\ldots$ is the principal value integral.
Here $\rho_{d}(\epsilon)$ is the impurity orbital density of states 
of the RLM
\begin{equation}
\rho_d(\epsilon)=\frac{1}{\pi}\,\frac{\Delta_{\rm eff}(\epsilon)}{
(\epsilon-g\mu_B h-\Lambda(\epsilon))^2+\Delta_{\rm eff}^2(\epsilon)} \ .
\end{equation}

The result (\ref{equ:gamma}) for $\gamma_{\rm imp}$ has some interesting
implications. First, because
$d^\dagger$ is connected to $S_+$, it is apparent that the low--energy excitations
in the system
are controlled by spin degrees of freedom, a well--known feature of the Kondo
physics. However,
in our approach this result can be read off directly from
equ.~(\ref{equ:gamma}). Second,
$\rho_{d}(0)\propto 1/\Delta_{\rm eff}(0)\propto 1/T_{\rm K}$, i.e.\ we obtain the
correct scaling behavior
for $\gamma_{\rm imp}$ {\em directly} from the behavior of
$\Delta_{\rm eff}(\epsilon)$. There is, however,
a nontrivial correction coming from the factor in parenthesis in
(\ref {equ:gamma}). Note that for $\Delta_{\rm eff}(\epsilon)=$const.\ this 
correction is one, but for the strongly $\epsilon$--dependent
$\Delta_{\rm eff}(\epsilon)$ in Fig.~\ref{fig:Fig1} it is of the order of two. As we
will demonstrate later,
this difference is directly responsible for obtaining the correct Wilson ratio
in our approach.

{}From the mapping $S_z = d^\dagger d-1/2$ it is easy to calculate
$\chi_{zz}(\omega+i\delta)
=-(g\mu_B)^2 \langle\langle d^\dagger d;d^\dagger d\rangle\rangle_{\omega+i\delta}$. Since
the correlation
function has to be evaluated within the RLM, one obtains for
the imaginary part
at $T=0$
\begin{equation}
  \label{equ:chipp}
  \chi_{zz}''(\omega)=(g\mu_B)^2 \:\pi  \int\limits_{-\omega}^0 d\omega'\:
\rho_{d}(\omega')\rho_{d}(\omega+\omega')\ .
\end{equation}
Again, this result provides direct access to an interpretation of the behavior
of $\chi_{zz}(\omega+i\delta)$
in terms of the physics of the resonant level model.

\section{Results}
\begin{figure}[htb]
\begin{center}
\includegraphics[width=0.45\textwidth,clip]{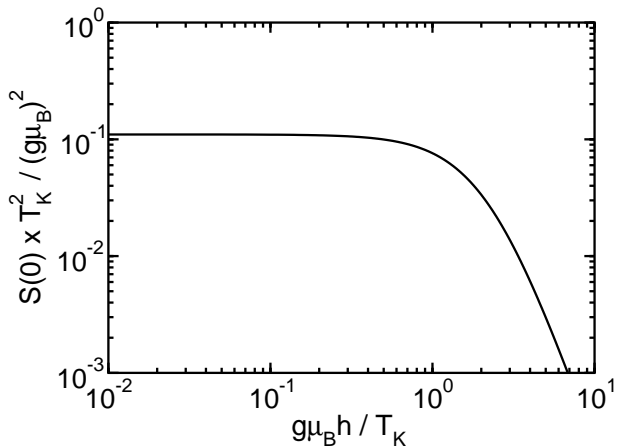}
\end{center}
\caption[]{Universal curve for $S(0)$ as function of the local magnetic field
$h$.\label{fig:Figt1}}
\end{figure}

One quantity that can be calculated analytically is the
low--energy limit of the spin
structure factor $S(\omega)\stackrel{\rm def}{=}\chi''_{zz}(\omega)/\omega$,
\begin{equation}\label{equ:Toneinvdef}
S(0)=\lim_{\omega\rightarrow 0}
\frac{\chi''_{zz}(\omega)}{\omega}\ .
\end{equation}
For a vanishing local magnetic field $S(0)$ is just the spin relaxation rate accessible
in e.g.\ spin resonance experiments.
With the result for $\chi_{zz}''(\omega)$ from (\ref{equ:chipp}) we obtain
\begin{equation}\label{equ:Toneinv}
S(0)=(g\mu_B)^2\:\frac{1}{\pi}\,
\frac{\Delta_{\rm eff}^2(0)}{\left((g\mu_B h)^2+\Delta_{\rm eff}^2(0)\right)^2}\;\; ,
\end{equation}
which leads to the curve shown in Fig.~\ref{fig:Figt1}.
Eq.~(\ref{equ:Toneinv}) is of particular importance because
% due to the relation $T_{\rm K}\propto \Delta_{\rm eff}(0)$
it explicitly demonstrates
universality, $T_{\rm K}^2S(\omega)=f(g\mu_B h/T_{\rm K})$, and allows to directly fit
e.g.\ experimental data from ESR or NMR experiments and extract the Kondo
temperature. Note furthermore that the result (\ref{equ:Toneinv}) is not
only valid in the Kondo limit, but also holds at the Toulouse point
of the anisotropic Kondo model and everywhere in between. Since it does not
depend on the details of $\Delta_{\rm eff}(\epsilon)$ it will also be true
for general band structures $\epsilon_{\vec k}$ in (\ref{equ:SIKM}) and thus
is eventually {\em the} result for $S(0)$ in DMFT calculations.

The full frequency dependent $\chi_{zz}''(\omega)$ has to be calculated numerically using the
form of the effective hybridization function in Fig.~\ref{fig:Fig1}.
The results for three values of the external field, $h=0$, 
$g\mu_B h=T_{\rm K}$ and $g\mu_B h=5T_{\rm K}$
\begin{figure}[htb]
\begin{center}
\includegraphics[width=0.45\textwidth,clip]{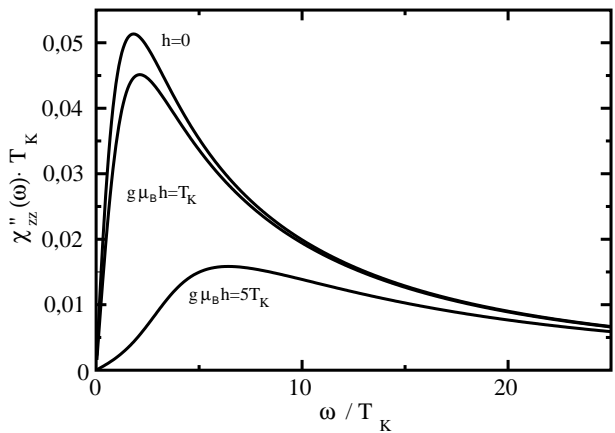}
\end{center}
\caption[]{$\chi_{zz}''(\omega)$ for three characteristic local magnetic fields
$h=0$, $g\mu_B h=T_{\rm K}$ and $g\mu_B h=5T_{\rm K}$.
\label{fig:chipp}}
\end{figure}
are displayed in Fig.~\ref{fig:chipp}. These correlation functions provide
a good example for
the usefulness of our mapping to the effective RLM since one can directly
interpret the structures and their frequency and field dependencies in terms of
analytical formulas derived for the RLM. E.g.\ the high--frequency behavior
of $\chi_{zz}''(\omega)$ follows directly from equ.~(\ref{equ:chipp})
and the behavior of the effective hybridization function
$\Delta_{\rm eff}(\epsilon)$ at large energies (which is linear with logarithmic
corrections, see Fig.~\ref{fig:Fig1}): $\chi_{zz}''(\omega)$ decays like
$1/\omega$ with logarithmic corrections, in agreement with (expensive)
numerical results \cite{Costi99}.

For the dependence of the dynamical susceptibility on the local magnetic field
one makes use of the fact that
the local magnetic field corresponds to the
on--site energy in the effective RML. Therefore it is obvious that the observed shift
of the resonance peak in $\chi_{zz}''(\omega)$ is due to the shifted center
of the resonant level. Furthermore, the depletion of the maximum value is
related to the decreasing occupation of the resonant level, which corresponds
directly to the {\em increasing} local magnetization in the SIKM. At the same
time, one observes a decrease of the total spectral weight in $\chi_{zz}''(\omega)$,
which can be accounted for by a transfer to a finite expectation value of
$\langle S_z\rangle$ in the SIKM. There is, however, also a non--trivial effect,
namely the increasing broadening of the resonance peak with increasing magnetic
field. For a RLM with a constant $\Delta_{\rm eff}(\epsilon)$ such a behavior does
not occur; it is entirely related to the fact that with increasing magnetic
field the system starts to notice the energy dependence of the effective
hybridization.

The quantity not yet fixed in our calculation is $T_{\rm K}$, or more
precisely the proportionality constant in $T_{\rm K}\propto \Delta_{\rm eff}(0)$.
This can most conveniently be done by using Wilson's definition
of the Kondo temperature\cite{hewson}
\begin{equation}
  \label{equ:ChiTk}
  \chi_0(h=0)=(g\mu_B)^2\,\frac{w}{4T_{\rm K}}\;\;,
\end{equation}
where $\chi_0$ is the static magnetic susceptibility
and $w=0.413$ the Wilson number. $\chi_0$
can be obtained from the imaginary part of the dynamic susceptibility
(\ref{equ:chipp}) via
\begin{equation}
  \label{equ:Chi}
  \chi_0=\frac{2}{\pi}\int\limits_0^\infty
d\omega\,\frac{\chi''_{zz}(\omega)}{\omega}
\end{equation}
and must in general be evaluated numerically.
At the Toulouse point one can, however, give an analytic answer since
$\Delta_{\rm eff}(\omega)={\rm const.}$ and thus
\begin{equation}
  \label{equ:ChiToulouse}
  \chi_0(h)=(g\mu_B)^2\:
\frac{1}{\pi}\,\frac{\Delta_{\rm eff}(0)}{h^2+\Delta_{\rm eff}(0)^2}\;\;.
\end{equation}
Therefore at the Toulouse point
the Korringa--Shiba relation\cite{KorringaShiba}
\begin{equation}\label{equ:KorringaShiba}
R_{\rm S}=\frac{(g\mu_{\rm B})^2}{2\pi\chi_0^2}\lim_{\omega\to0}
\frac{\chi_{zz}''(\omega)}{\omega}
\end{equation}
is independent of the local magnetic field
\begin{equation}
  \label{equ:ShibaToulouse}
R_{\rm S}=  \frac{(g\mu_B)^2\,S(0)}{2\pi\chi_0^2(h)}=\frac{1}{2} \ .
\end{equation}

In the following we will discuss $\chi_0(h)$ and the Korringa--Shiba relation
for the Kondo limit $\rho_0 J \rightarrow 0$.
The quantity $\chi_0(h)$ is particularly convenient for a comparison with NRG results.
\begin{figure}[htb]
\begin{center}\mbox{}
\includegraphics[width=0.45\textwidth,clip]{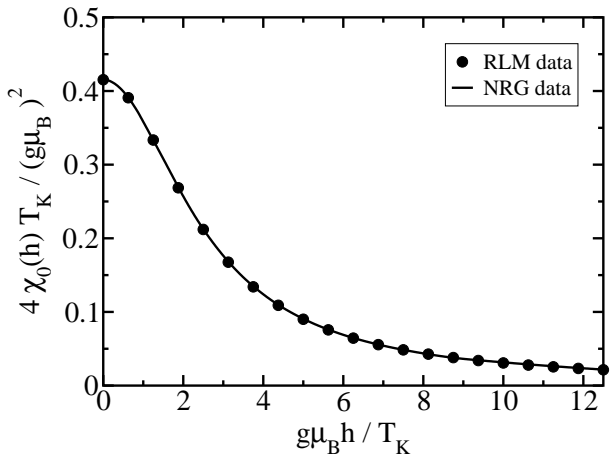}
\end{center}
\caption[]{The magnetic susceptibility $\chi_0(h)$ from equ.~(\ref{equ:Chi}) (circles)
and the same quantity obtained from an NRG calculation.\label{fig:Fig2}}
\end{figure}
In Fig.~\ref{fig:Fig2} the circles represent the values of $\chi_0(h)$
calculated via (\ref{equ:Chi}) with the effective hybridization function
from Fig.~\ref{fig:Fig1}, and the full line represents the result of an NRG calculation.
We observe excellent agreement for all values of the local magnetic field:
notice that the curves agree without fit parameters.
This example clearly demonstrates
that the nontrivial form of the effective hybridization in
Fig.~\ref{fig:Fig1} encodes the many--particle physics of the SIKM
in a trivial noninteracting effective model.

\begin{figure}[htb]
\begin{center}\vspace*{5mm}
\includegraphics[width=0.45\textwidth,clip]{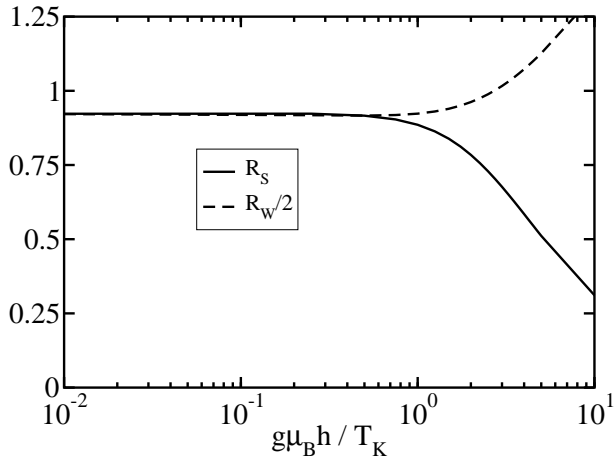}
\end{center}
\caption[]{Results for the Shiba ratio $R_{\rm S}$ (full line) and
the Wilson ratio $R_{\rm W}$ (dashed line) as a function of a local magnetic
field. The
correct limiting values at $h\to0$ are missed by approximately 5\%.
\label{fig:Fig3}}
\end{figure}

The result in Fig.~\ref{fig:Fig2} can readily be combined with relations
(\ref{equ:gamma}) and (\ref{equ:Toneinv}) to obtain the Wilson ratio\cite{Wilson}
\begin{equation}\label{equ:Wilson}
R_{\rm W}=\frac{4\pi^2k_{\rm B}^2}{3(g\mu_{\rm B})^2}
\frac{\chi_{\rm imp}(h)}{\gamma_{\rm imp}(h)} \ .
\end{equation}
Our results for the Wilson ratio and the Korringa--Shiba relation obtained
within the effective RLM are collected
in Fig.~\ref{fig:Fig3}.
For the Wilson ratio we would actually have to calculate the quantity
$\chi_{\rm imp}$  and not $\chi_0$.\cite{Wilson}
However, for the case of small $\rho_0J$ considered here, both
quantities are equivalent.\cite{chen}
One observes that both $R_{\rm W}$ and $R_{\rm S}$ are independent of the local magnetic
field up to approximately $g\mu_B h\approx T_{\rm K}$, and then start to
decrease (Shiba ratio), respectively increase (Wilson ratio).
The exact Bethe ansatz solution \cite{bethe} gives $R_{\rm W}(h)=2$
independent of the magnetic field strength (see also
Ref.~\onlinecite{logan3}), and local Fermi liquid theory yields
$R_{\rm S}=1$ for $h\rightarrow 0$.\cite{hewson}
Our limiting values as $h\to0$ miss these
exact results by approximately 5\%.
Notice that the term $(1-\Lambda'(0))$ in (\ref{equ:gamma})
is very important to obtain this correct value for $R_{\rm W}(h=0)$.
Remarkably, our simple {\em noninteracting}
effective model therefore correctly describes the Wilson ratio in the
Kondo limit (for not too large magnetic fields), which is a hallmark of
strong--coupling Kondo physics.

Let us finally analyze the accuracy of our effective model.
Since Fig.~\ref{fig:Fig2} demonstrates
that integral quantities like $\chi_0(h)$ are obtained with very good
accuracy for {\em all} magnetic fields, one can infer from Fig.~\ref{fig:Fig3}
that quantities depending on low--energy details in frequency
space like $\gamma_{\rm imp}$ and $S(0)$ are more susceptible to our
approximations for increasing magnetic fields.
This suggests that for such low--energy quantities our effective model
can be employed with very good accuracy (5\%~error) for magnetic fields below
$T_{\rm K}$, and with good accuracy (20\%~error) still up to approximately~$5 T_{\rm K}$.

\section{Summary and outlook}
Summing up, we have shown that the resonant level
model with a nontrivial hybridization function $\Delta_{\rm eff}(\epsilon)$
can be used as an effective model
for the single impurity Kondo model. The key observation was the fact
that the flow equation solutions of both models are approximately
identical if a suitable $\Delta_{\rm eff}(\epsilon)$ is chosen for the RLM (Fig.~1).
In this mapping the impurity orbital occupation number $n_d-1/2$ plays the
role of the Kondo impurity spin $S_z$. The impurity orbital energy
corresponds to the local magnetic field acting on~$S_z$.

In contrast with the conventional approach where effective models
describe the vicinity of the low--energy renormalization group fixed
points\cite{Wilson,hewson}, our effective model very accurately describes {\em both
certain low-- and high--energy
properties of the original Kondo model}: compare for example
our discussion of the dynamical spin--spin correlation function
in Fig.~\ref{fig:chipp}. It also yields
thermodynamic quantities that are in excellent agreement with
much more expensive numerical methods (see Fig.~\ref{fig:Fig2}). The nontrivial
behavior of the effective hybridization function encodes the
quasiparticle interaction, which leads to e.g.\ the correct Wilson
ratio for small magnetic fields (with 5\% accuracy). Notice,
however, that our effective model does
not allow the correct evaluation of higher--order correlation functions
beyond the low--energy limit,

In conclusion, our approach describes many aspects of the complicated many--body
Kondo physics for not too large magnetic fields
within a simple noninteracting model. Therefore one can very
easily and intuitively understand certain properties of the Kondo
model, e.g. the dependence of correlation functions on a local magnetic
field (Fig.~\ref{fig:chipp}). One main prospect of our approach is to look at
other correlation functions using this effective model, in particular
the $T$--matrix for applications in the framework of DMFT calculations.
Future prospects also include cluster problems and the single impurity
Anderson model.
Work along these lines is in progress.
\begin{acknowledgments}
We acknowledge valuable conversations with
N.~Andrei, R.~Bulla, W.~Hofstetter, D.~Logan, A.~Rosch, M.~Vojta and
D.~Vollhardt.
This work was supported by the DFG collaborative research center SFB~484 and NSF
grant DMR-0073308.
\end{acknowledgments}


\begin{thebibliography}{99}
\bibitem{perspective} The investigation of nontrivial dispersion relations is
one of the main future perspectives of the flow equation approach. Work
along these lines is in progress.
\bibitem{numerics}
It becomes increasingly difficult to numerically investigate smaller couplings.
\end{thebibliography}
\end{document}